\documentclass{article}
\usepackage{authblk}
\usepackage{graphicx}
\usepackage{subfigure}
\usepackage{amsmath}
\begin{document}
\title{Quantization of B-I electrodynamics and B-I modified gravity using Faddeev-Popov gauge-fixing procedure}
\author{Srivatsan Rajagopal\thanks{chichieinstein@sify.com}\\Ajit Kumar\thanks{ajitk@physics.iitd.ac.in}}
\affil{Department of Physics, Indian Institute of Technology\\
       Hauz Khas, New Delhi-110016}
\maketitle
\begin{abstract}
We investigate the quantized versions of Born Infeld electrodynamics and Born Infeld Gravity. We derive Feynman rules for B-I electrodynamics by deriving an effective Lagrangian with the square root removed using the Faddeev-Popov method. In the case of B-I gravity, the square root in the Lagrangian is removed by the introduction of the Vierbien fields. This approach has the advantage that SO(3,1) can be consistently regarded to be the gauge group of gravity. Finally, using a rough argument, the quantum fluctuations of the radii of spatial hypersurfaces in flat space are shown to undergo accelerated increase with time.
\end{abstract}

\newpage
\section*{Introduction}
Recently, in order to account for the accelerated expansion of the universe, suitable modification of Einstein's field equation of general relativity has become a hot topic of research\cite{SC1,SC2}. Such modified theories are now called f(R) gravity. In this context various modified models with higher-order derivatives have been put forward and rigorously investigated. Although there are problems with the quantization and unitarity of in f(R) gravity, it is expected to resolve
the coincidence problem, explain the existence of dark matter and account for the transition from deceleration to acceleration \cite{Noj,Cap,Kru,Wood,Cap2,Pan,Del}. One of the variations is to generalize Einstein's equation by adding terms similar to those proposed by Born and Infeld for Maxwell's electrodynamics\cite{Born}. To the best of
our knowledge, in the context of gravity, Deser and Gibbon\cite{Deser} were the first to study the viability of Einstein-Born-Infeld action in four dimensions. At present,given the recent developments in string theory and cosmology, Born and Infeld (B-I) nonlinear electromagnetic field theory has been experiencing an astonishing renaissance\cite{SMC,JCF,Come,Padma,Sami,Gib2,Des2,Banados}. The Born-Infeld action, also referred to as Dirac-Born-Infeld action, arises as the low- energy effective action of open strings, and as part of the world-volume effective action of D3-branes\cite{Ket,Princeton}.\\

The $\sigma$-model for the world-sheet of the string was shown to require the background field to be described by B-I action[20, 21]. Other particularly attractive aspects of the B-I action in electrodynamics are its famous taming of the Coloumb selfenergy divergence, and the absence of birefringence. Currently, B-I theory has been used in various branches of physics including non-liner optics, condense matter physics, nuclear physics, high-energy physics, string theory and Cosmology. \\

In view of these interesting characteristics, we have attempted here to study some of its properties. We have earlier\cite{us} showed that the Ostrogradsky instability does not plague Born-Infeld gravity when it is varied in the Palatini
formalism. Our aim here is to investigate quantized B-I electrodynamics and B-I modified gravity.\\

The principal difficulty in quantizing this theory is caused by the appearance of a square root in the Lagrangian. In B-I Electrodynamics, this is removed by the application of the Faddeev-Popov method and an effective Lagrangian is derived. In the case of gravity, there is a more natural procedure to remove the square root. This is done by the introduction of Vielbien (or Tetrad) fields. The gravitational connection then becomes a gauge field. An important point of consistency
is that throughout our work, we have been using the Palatini formalism where the metric and connection are independent. Hence, in the language of tetrad fields, the Vielbien (which is determined solely by the metric) and the connection (which is the gauge field) become independent dynamical variables \cite{Knots,Moshe}. Here, we have chosen to concentrate on the gauge field and have taken the tetrads to be those of Minkowski space where it is easier to do quantum field theory.

\section*{Born Infeld Electrodynamics}

The Lagrangian density for the B-I field is given by:
\begin{align}
L = (1+ F_{\mu\nu}F^{\mu \nu}/\beta^2 - (F_{\mu \nu}B^{\mu \nu})^2/\beta^4)^{1/2}
\end{align}
Here, as usual, $F^{\mu \nu}$ represents the Electromagnetic field strength tensor while $B^{\mu \nu} = \epsilon^{\mu \nu \alpha \beta}F_{\alpha \beta}$ represents its dual. Following \cite{Euler}, in order to remove the square root, we introduce an auxiliary field $\xi$ as follows 
\begin{align}
L = \frac{-\xi}{2} \left(1+ \frac{F_{\mu\nu}B^{\mu \nu}}{\beta^2} - \frac{(F_{\mu \nu}B^{\mu \nu})^2}{\beta^4} \right) - \frac{1}{2\xi}
\end{align}
We observe that the $\xi$ field enters the Lagrangian without time derivatives and hence, it has no independent dynamics. When eliminated between its equations of motion and the equations for the $A_\mu$, the same equations as the original Lagrangian are obtained. The advantage of introducing this field is that on passing to the path integral, $\xi$ can be
taken to be independent of the other fields, since in the path integral, the action is not stationary but acquires all values with same probability. The Faddeev-Popov procedure is then used here, and consists of exploiting the gauge
invariance to factor out the infinite volume of the Gauge space as discussed. For the Abelian case of electrodynamics, this throws in a factor that is independent of the $A_\mu$ and is not important for our approach. After doing all that, the
normal course \cite{Ryder} is to insert an arbitrary field $c$ and averaging over all such fields. In our approach, we choose as $c$ the extra auxiliary field which we have introduced in the Lagrangian above. The validity of this choice can be justified as follows.\\

The path integral is given by:
\begin{align}
Z = \int D\xi DA_\mu e^{iS(\xi , A_\nu)}
\end{align}
Here, $S$ represents the action which depends on the dynamical variables $\xi$ and $A_{\mu}$. Since $\xi$ and $A_\mu$ are independent degrees of freedom, we can proceed with the standard Faddeev-Popov integration for $A_\mu$ to get
\begin{align}
Z = \int D\xi DA_\mu \Delta _F[A_\mu] \delta(g(A_\mu)-c) e^{iS(\xi , A_\nu)}
\end{align}
Here, $\Delta_F(A_\mu)$ is the term arising out of neglecting the infinite volume element of the gauge group space. For Abelian gauge theory, it is independent of $A_{\mu}$ and can be taken out of the integral. Also, $g(A_\mu)$ is the gauge fixing condition that one would like to impose.\\

The important point to observe here is that the arbitrary field $c$ introduced above is completely independent of $A_\nu$ . Therefore, it can be taken to be $\xi$ itself. The justification for this is that there is gauge invariance in the equations. This freedom can be used to set the gauge function $g(A_\mu)$ equal to any arbitrary function. Explicitly, if $A'_\mu = A_\mu + \partial _\mu \lambda$, so that, if $\partial_\mu A^\mu = 0$ then, $g(A'_\mu) = \partial_\mu ^\mu \lambda = \xi$ which can be solved for $\lambda$. Thus in the path integral, the gauge fixing function can be set equal to $\xi$.\\

The main point of difference is that instead of the traditional gauge breaking term in the effective Lagrangian, which is quadratic in the $g(A_\nu)$, we get the following effective Lagrangian after integrating over the $\xi$.
\begin{align}
L_{eff} = -\frac{g(A_\nu)}{2}\left(1+ \frac{F_{\mu \nu}F^{\mu \nu}}{\beta ^2}- \frac{(F^{\mu \nu} B_{\mu \nu})^2}{\beta ^4}\right) - \frac{1}{2g(A_\nu)}
\end{align}

Finally, since, as we said above, the auxiliary field $\xi$ is not dynamical, we can restrict to variations of the Lagrangian such that $\delta(g(A_\mu))$ = 0. Therefore, the first and last terms
can be neglected to get
\begin{align}
L_{eff} = \frac{g(A_\mu)}{2}\left(\frac{F_{\mu \nu}F^{\mu \nu}}{\beta ^2} - \frac{(F^{\mu \nu} B_{\mu \nu})^2}{\beta ^4} \right)
\end{align}

Hence, we have achieved our aim: we have removed the square root and brought the Lagrangian to polynomial form in the $A_\mu$ and their derivatives. Now, we come to the problem of getting the Feynman rules for this effective Lagrangian.

\section*{The Feynman rules for Born Infeld Electrodynamics}

We can now proceed in the usual way, \cite{Ryder} finding the Feynman rules for B-I electrodynamics using functional differentiation in momentum space. The action can be written as the sum of a third order and a fifth order term as (after partial integrations) :
\begin{align}
S_{eff} = -\int A^{\alpha} \left( \frac{(F_{\mu \nu}B^{\mu \nu})\partial _{\alpha}(F_{\mu \nu}B^{\mu \nu})}{\beta ^4} + \frac{\partial _\alpha F_{\mu \nu}F^{\mu \nu}}{\beta ^2}\right) d^4 x
\end{align}

Here, we would like to define a few symbols before proceeding further; we define $\epsilon ^{\mu \nu} _ {\mu_1 \nu_1}$ to represent the quantity that is +1 for ($\mu_1 \nu_1$) = ($\mu \nu$) and -1 for the cyclic permutation. $\epsilon_{\alpha \beta \gamma \delta}$ is the usual permutation tensor defined for four indices. With these
definitions, the action above becomes

\begin{align}
S_{eff} = -\int \frac{A^\alpha}{\beta ^4} \left(\epsilon_{\mu \nu \gamma \delta} F^{\mu \nu } F^{\gamma \delta} \right) \partial _\alpha \left(\epsilon_{\mu _1\nu _1 \gamma \delta} F^{\mu_1\nu_1} F^{\gamma \delta}\right) + \frac{A^\alpha}{\beta^2}\left(F^{\mu \nu} \partial _{\alpha} F_{\mu \nu}\right) d^4 x
\end{align}

Renaming a few indices and expanding, we get the integrand
\begin{align}
\int \frac{A^{\alpha}}{\beta^4}\left(\epsilon_{\mu \nu \gamma \delta} \epsilon ^{\mu \nu}_{\mu_1 \nu_1} \epsilon^{\gamma \delta}_{\sigma_1 \rho_1 } \partial ^{\mu_1}A^{\nu_1} \partial^{\sigma_1}A^{\rho_1}\right) \partial_{\alpha} \left(\epsilon^{\mu_1 \nu_1}_{\eta \theta} \epsilon^{\gamma \delta}_{\gamma_1 \delta_1}\partial ^{\eta} A^{\theta}\partial^{\gamma_1}A^{\delta^1}\right) d^4 x   +\nonumber \\* \int \frac{A^\alpha}{\beta^2}\left(\epsilon^{\mu \nu}_{\mu_1 \nu_1 }\epsilon^{\alpha_1 \beta_1}_{\mu \nu}\partial^{\mu_1}A^{\nu_1}\partial _{\alpha} \partial_{\alpha_1} A_{\beta_1} \right) d^4 x
\end{align}

The term giving the dominant contribution in perturbation theory is the second term. Therefore, we now concentrate on this term. We first raise the last $\beta_1$ index and write
\begin{align}
S_3 = \int \frac{A^{\alpha}}{\beta^2} \eta_{\beta_1 \gamma} \left(\epsilon^{\mu \nu}_{\mu_1 \nu_1 }\epsilon^{\alpha_1 \beta_1}_{\mu \nu}\partial^{\mu_1}A^{\nu_1}\partial_{\alpha} \partial_{\alpha_1} A^{\gamma} \right) 
\end{align}
Here, $\eta_{\alpha \beta}$ is the Minkowski metric tensor.
To find the functional derivatives in momentum space, we first use the Fourier transformation as usual and write
\begin{align}
A^{\alpha}(x) = \int A^{\alpha}(p) e^{ip.x} d^4 p \nonumber \\
A^{\nu_1}(x) = \int A^{\nu_1}(q) e^{iq.x} d^4 q  \nonumber \\
A^{\gamma}(x) = \int A^{\gamma}(r) e^{ir.x} d^4 r  
\end{align}

Substituting this in (9), we get 
\begin{align}
S_3 = \frac{-i}{\beta^2} \eta_{\beta_1 \gamma} \int \delta(p+q+r) A^{\alpha}(p)A^{\nu_1}(q)A^{\gamma}(r)\left(\epsilon^{\mu \nu}_{\mu_1 \nu_1 }\epsilon^{\alpha_1 \beta_1}_{\mu \nu}\right) q^{\nu_1} r_{\alpha} r_{\alpha_1} d^4 q d^4 p d^4 r
\end{align}

The delta function arises when the integration over the position coordinates is carried out.
The functional derivatives are now straightforward to compute and following \cite{Ryder} are given by 
\begin{eqnarray}
&&\frac{\delta^3 S_{eff}}{\delta A^{\rho}(p) \delta A^{\sigma}(q) \delta A^{\lambda}(r)} =\nonumber\\
&&-\frac{i}{\beta^2}\delta(p+q+r) \left(\eta_{\beta\lambda}\epsilon^{\mu\nu}_{\mu_1\sigma}\epsilon^{\alpha\beta}_{\mu\nu}q^{\mu_1}r_{\rho}r_{\alpha}+ \eta_{\beta\sigma}\epsilon^{\mu\nu}_{\mu_1\lambda}\epsilon^{\alpha\beta}_{\mu\nu}r^{\mu_1}q_{\rho}q_{\alpha}\right)\nonumber\\
\nonumber\\
&&\hskip -0.5cm -\frac{i}{\beta^2}\delta(p+q+r) \left(\eta_{\beta\lambda}\epsilon^{\mu\nu}_{\mu_1\rho}\epsilon^{\alpha\beta}_{\mu\nu}p^{\mu_1}r_{\sigma}r_{\alpha}+\eta_{\beta\sigma}\epsilon^{\mu\nu}_{\mu_1\rho}\epsilon^{\alpha\beta}_{\mu\nu}p^{\mu_1}q_{\lambda}r_{\alpha}\right) \nonumber\\
\nonumber\\
&&\hskip -0.5cm -\frac{i}{\beta^2}\delta(p+q+r) \left(\eta_{\beta\rho}\epsilon^{\mu\nu}_{\mu_1\lambda}\epsilon^{\alpha\beta}_{\mu\nu}r^{\mu_1}p_{\rho}p_{\alpha}+\eta_{\beta\rho}\epsilon^{\mu\nu}_{\mu_1\sigma}\epsilon^{\alpha\beta}_{\mu\nu}q^{\mu_1}p_{\lambda}r_{\alpha}\right).
\end{eqnarray}

\section*{Quantised Born-Infeld Gravity}

In this part, we would focus on the quantisation of Born-Infeld gravity. Specifically, we would like to show that a part of the Lagrangian can be written completely in analogy with Yang-Mills Lagrangian. The procedure to do that is as follows:

\begin{eqnarray}
&&\hskip -1.0cm S =\lambda^4 \int \text{det}(e^\mu_a) d^4 x  + \lambda^3 \int R \text{det}(e^\mu_a) d^4 x+ \frac{\lambda^2}{2!} \int \text{det}(e^\mu_a) \left( R^2 - R^{\mu \nu}R_{\mu \nu}\right) d^4 x \nonumber\\ 
&+& \frac{\lambda}{3!} \int \left(R^3 - 3RR^{\mu \nu}R_{\mu \nu} + 2R^{\mu \alpha}R_{\alpha \beta}R^{\beta}_{\mu}\right) \text{det}(e^\mu_a) d^4 x \nonumber\\
&+& \int \text{det}(e^\mu_a) \text{det}(R_{\mu\nu})d^4 x    
\end{eqnarray}

Because of the smallness of the cosmological constant, the last two terms can be neglected. Moreover, we are using the Palatini formalism. Therefore, the Vielbien and the Connection field are take as independent dynamical variables. We choose here a background of Minkowski space and only proceed to quantise the Connection field. This implies that the first term is non-dynamical. Moreover, we are interested only in the two point correlation functions. Therefore, the first term is uninteresting in our case.

The interesting term is therefore given by 
\begin{align}
S = \frac{\lambda^2}{2} \int(R^2 - R_{\mu \nu}R^{\mu \nu}) d^4 x
\end{align}

where $e = det(e^\mu_a) = 1$ because the tetrads are those of flat spacetime.
The term $R^{\mu\nu} R_{\mu\nu}$ can be written as 
\begin{align}
R^{\mu \nu}R_{\mu \nu} = e^{\sigma}_a e^{\gamma}_c F^{a\rho}_{d\sigma} F^{cd}_{\rho \gamma}
\end{align}

The product of tetrads can be decomposed as 
\begin{align}
e^{\sigma}_a e^{\gamma}_c = \frac{1}{4} g^{\sigma \gamma}\eta_{ac} + A^{\sigma \gamma}_{ac}
\end{align}
where $A^{\sigma \gamma}_{ac}$ satisfies $g_{\sigma \gamma} A^{\sigma \gamma}_{ac} = 0$

With this decomposition, the $R^{\mu \nu}R_{\mu \nu}$ term can be written as
\begin{align}
R^{\mu \nu} R_{\mu \nu} = \frac{1}{4}g^{\sigma \gamma}\eta_{ac} F^{a\rho}_{d\sigma} F^{cd}_{\rho \gamma} + A^{\sigma \gamma}_{ac}F^{a\rho}_{d\gamma} F^{cd}_{\rho \sigma}
\end{align}

the RHS of which is

\begin{align}
\frac{1}{4} F^{a\rho}_{d\sigma} F^{d\sigma}_{a\rho} + A^{\sigma \gamma}_{ac}F^{a\rho}_{d\gamma} F^{cd}_{\rho \sigma}
\end{align}
and hence the first term can be clearly seen to be the double trace of $F^{\mu \nu}F_{\mu \nu}$ and can be written as
\begin{align}
\frac{1}{4} Tr(F^{\mu \nu}F_{\mu \nu}) 
\end{align}
The first part can be clearly seen to be analogous to the kinetic part of the Yang-Mills field. Here the $Tr$ signifies a double trace and hence the analogy is not totally complete.

\section*{Feynman Rules for Quantized B-I Gravity}

The kinetic part can be used to get the propagator in the usual way \cite{Ryder}. This is given by 

\begin{align}
D^{abcd}_F(x-y) = -i\delta^{ab}\delta^{cd} \int g_{\mu\nu} \frac{e^{-ik.(x-y)}}{k^2 + i\epsilon} d^4 k.
\end{align}

Note that the $g_{\mu \nu}$ is now the flat space Minkowski metric as argued above.\\

Finally, following the method used in deriving (12), we calculate the term in the three vertex rule, arising from the kinetic part, as given below
\begin{eqnarray}
G(p,q,r)&=&\frac{\delta^3 S_{eff}}{\delta A^{\alpha}_{a_1b_1}(p) \delta A^{\beta}_{c_1d_1}(q) \delta A^{\gamma}_{e_1f_1}(r)}= \nonumber\\
\nonumber\\
&-&i \delta(p + q + r)\left(p_\beta g_{\gamma \alpha}-p_{\gamma}g_{\beta \alpha}\right)\eta^{aa_1}\eta^{bb_1}C^{c_1d_1e_1f_1}_{ab}\nonumber\\
\nonumber\\
&-&i \delta(p + q + r)\left(q_{\alpha}g_{\beta \gamma}-q_{\gamma} g_{\alpha \beta}\right) \eta^{ac_1}\eta^{bd_1}C^{a_1b_1e_1f_1}_{ab}\nonumber\\
\nonumber\\
&-& i \delta(p+q+r)\left(r_\gamma g_{\beta \alpha}- r_\beta g_{\alpha \gamma}\right)\eta^{ae_1}\eta^{bf_1}C^{a_1b_1c_1d_1}_{ab}.
\end{eqnarray}
Here $C^{abcd}_{ef}$ are the structure constants of the SO(3,1) group.\\

\section*{A heuristic argument for the curvature fluctuations of spatial hypersurfaces}

Finally, we show that the kinetic energy part of the Lagrangian in (20) results in an accelerated increase in the quantum fluctuations of spatial hypersurfaces. This is shown by first considering the propagator derived above. The two point correlation function of the same components of the spatial Gauge field at a local point as a function of time are given by

\begin{align}
\langle A_i(0) A_i(t) \rangle = D_F(t) = -i\int \frac{e^{ik_4t}}{k^2-k_4^2} d^4 k.
\end{align}

Using the analyticity of the path integral, we perform analytic continuation to imaginary time and replace $t = i\tau$. Also, we use Wick rotation of the contour along which (18) is evaluated. Under these transformations, the integral in (18) is evaluated to
\begin{align}
D_F(\tau) = -\frac{K}{4\pi \tau^2} = \frac{K}{4\pi t^2},
\end{align}

where, again by analytic continuation, we have replaced τ by – it. Here, K is a positive constant. Then, $D_F(t)$ is (roughly) proportional to the fluctuations in the extrinsic curvature of the spatial hypersurfaces. The fluctuation in radius can be roughly taken to be the reciprocal of (19). As can clearly be seen, this fluctuation has a positive second derivative.\\

\section*{Conclusions}

In this paper, we have explored the possible quantisation procedures for both Born-Infeld gravity and electromagnetism. In the case of Born-Infeld electromagnetism, the Faddeev-Popov method was applied to remove the square root and to find the Feynman rules for 3 vertex interactions in this theory. Though our treatment of gravity as a gauge field is not new, we have used the Palatini formalism to our advantage to focus mainly on the connection part of the action which has become the sole dynamical quantity in our procedure. This has some appeal since the it is the connection field that carries the true signature of gravity.

\end{document}